\documentclass[twocolumn]{jpsj2} 
\usepackage{dcolumn}
\usepackage{bm}
\usepackage{amsmath}
\usepackage{color}

\title{A Numerical Model for Brownian Particles Fluctuating in Incompressible Fluids}

\author{Takuya \textsc{Iwashita}$^{1}$\thanks{E-mail: iwashita@cheme.kyoto-u.ac.jp}, Yasuya \textsc{Nakayama}$^{2}$, and Ryoichi \textsc{Yamamoto}$^{1,3}$}

\inst{$^{1}$Department of Chemical Engineering, Kyoto University, Kyoto 615-8510 \\
$^{2}$Department of Chemical Engineering, Kyushu University, Fukuoka 819-0395 \\
$^{3}$ CREST, Japan Science and Technology Agency, Saitama 332-0012}

\abst{We present a numerical method that consistently implements thermal fluctuations and hydrodynamic interactions to the motion of Brownian particles dispersed in incompressible host fluids. 
In this method, the thermal fluctuations are introduced as random forces acting on the Brownian particles. 
The hydrodynamic interactions are introduced by directly resolving the fluid motions with the particle motion as a boundary condition to be satisfied. 
The validity of the method has been examined carefully by comparing the
 present numerical results with the fluctuation-dissipation theorem
 whose analytical form is known for dispersions of a single spherical 
particle. 
Simulations are then performed for more complicated systems, such as
a dispersion composed of many spherical particles and a single polymeric
 chain in a solvent.}

\kword{Brownian motion, hydrodynamic interaction, thermal fluctuation, particle dispersions, direct numerical simulation, long-time tails, polymer dynamics}

\begin{document}
\maketitle

\section{Introduction} 
The dynamics of solid particles dispersed in host fluids is very
complicated. 
Although computer simulations have been extensively used as a tool to
investigate those systems, obtaining meaningful results is not yet
straightforward even for the simplest case where the particles are
monodisperse spheres and the host fluid is Newtonian.
The main difficulty comes from the consistent treatment of the so-called
hydrodynamic interaction (HI) between particles mediated by fluid
motions, which are induced by the particle's motion.

The mathematical expression for HI is greatly simplified if the
following assumptions are made: 1) the dispersed particles are all
spherical; 2) HI acting among particles is pair-wise additive; and 3)
the motions of the host fluid instantaneously follows the motions of the
particles (the Stokes approximation). 
Then, the HI is expressed as a tensor that is a function of the
particle's positions and velocities, without explicitly dealing with the
fluid motions.  
The Oseen and the Rotne-Prager-Yamakawa (RPY) tensors are probably the
most well known forms of such simplified HI functions; the former
neglects the size of the particle, while the latter takes into account
some corrections to the particle size. 
The Stokesian dynamics (SD) method \cite{citeulike:2365048} is a widely
used numerical method along the lines of solving the tensor equations. 
It is based on the Langevin-type equations for particles implementing
the RPY tensor and the lubrication correction. 
The latter is necessary when the distance between particles is small
compared to the particle radius. 

Completely different numerical approaches have been developed recently
\cite{LB1,LB,SR,NS,NS1}, where the motions of host fluids are explicitly
solved along with the motion of the particles so that the HI is directly
computed without using the three assumptions described above.
We refer those approaches as direct numerical simulation (DNS)
approaches.
An apparent benefit of using DNS approaches for particle dispersions is
that the long-time behavior of particle motions is reproduced accurately.
For example, the velocity auto correlation function (VACF) of a
fluctuating particle is expected to show a non-exponential power-law
relaxation-- known as the hydrodynamic long-time tail--if the thermal
fluctuations are taken into account as well. 

Recently, we also proposed an efficient DNS scheme for particle
dispersions called the Smoothed Profile (SP) method\cite{nakayama:036707, Naka}, where the
discontinuous boundaries between particles and a fluid are smoothed out
by using a continuous profile function, thereby achieving greater
computational efficiency. 
The SP method has been successfully applied to some problems, such as
the electrophoresis of charged colloidal particles\cite{Kim}; however, the effects
of thermal fluctuations (TF) are neglected. 
This is not a bad approximation for large/heavy particles, but it is
insufficient for particles whose size is on the order of a micrometer or
smaller; the coupling of TF and HI becomes crucial in these systems. 
Owing to some new experimental techniques that enable direct examination
of the properties of Brownian particles fluctuating in a host fluid,
several interesting phenomena have recently been reported, including the
non-diffusive behavior of Brownian particles \cite{luki:160601} and the
rotational-translation coupling of a pair of spherical particles
\cite{martin:248301} where the coupling of TF and HI plays an essential
role. 

In the present study, our main goal is to implement TF and HI
consistently within a DNS framework for particle dispersions. 
We aim to introduce TF to our original SP method to achieve this end.
Since theoretical analyses have been well established for dilute
dispersions, we first perform simulations for a dilute dispersion
composed of a single spherical particle in a cubic box with the periodic
boundary condition. 
The numerical results for the VACF are then compared with analytical
solutions based on the fluctuation-dissipation theorem.
After the validity of the method has been confirmed for this simple
system, simulations are performed for dense dispersions composed of 
many spherical particles for which analytical solutions are unknown.
We furthermore performed DNS simulations to examine the dynamics of a single
polymeric chain fluctuating in a good solvent. 
The time-correlation functions are calculated for each Rouse mode of the
chain and compared with the predictions of the Rouse and Zimm models.

\section{Simulation Method}

We now present our implementation of TF based on the Langevin type approach.
The equation governing the solvent with a density $\rho_f$ and a shear
viscosity $\eta$ is a modified Navier-Stokes equation
\begin{eqnarray}
\rho_f(\partial_t {\bm v} + {\bm v}\cdot \nabla {\bm v})=-\nabla p
 +\eta\nabla^2 {\bm v}+\rho_f\phi{\bm f_p}
\label{nseq}
\end{eqnarray}
with the incompressible condition $\nabla\cdot {\bm v}=0$, where ${\bm
v({\bm x},t)} ,p({\bm x},t)$ is the velocity and pressure field of the
solvent. 
{A smooth profile function $0\leq \phi({\bm x}, t)\leq 1$
distinguishes between fluid and particle domains, namely
$\phi=1$ in the particle domain and $\phi=0$ in the fluid domain. Those
domains are separated by thin interfacial regions whose thickness is
characterized by $\xi$. 
The body force $\phi {\bm f_p}$ is introdued to ensures the
rigidity of particles and the non-slip appropriate boundary condition
at the fluid/particle interface.
The mathematical expressions of $\phi$ and $\phi {\bm f_p}$ are given 
in earlier papers \cite{nakayama:036707, Naka} in detail.}

We consider dispersions composed of $N_p$ spherical particles with a radius $a$.
The motion of the $i$th particle is governed by Newton's equations of motion
with stochastic forces: 
\begin{align}
M_i \dot {\bm V_i}&= {\bm F^H_i} + {\bm F^C_i} + {\bm G_i^V},\ \ \
\dot {\bm R_i} = {\bm V_i},\\
{\bm I_i}\cdot \dot{\bm \Omega_i} &= {\bm N^H_i} + {\bm G_i^\Omega},
\end{align}
where ${\bm R_i}$, ${\bm V_i}$, and ${\bm \Omega_i}$ are the position,
translational velocity, and rotational velocity of particles,
respectively.
$M_i$ and ${\bm I_i}$ are the mass and the moment of inertia, 
and ${\bm F^H_i}$ and ${\bm N_i^H}$ are the hydrodynamic force and torque
exerted by solvent on the particle.\cite{nakayama:036707, Naka}
${\bm F^C_i}$ is direct interparticle interaction such as Coulombic or
the Lennard-Jones potential. In the present study, we used the truncated
Lennard-Jones interaction: 
\begin{align}
U_{LJ}(r_{ij}) &=
\begin{cases}
 4\epsilon\Bigl[\Bigl(\frac{\sigma}{r_{ij}}\Bigl)^{12} -
 \Bigl(\frac{\sigma}{r_{ij}}\Bigl)^{6}\Bigl)\Bigl] + \epsilon & 
 (r_{ij} <2^{1/6}\sigma),\\
 0 & (r_{ij}>2^{1/6}\sigma)
\end{cases}
\end{align}
{where $r_{ij}=|\bm R_i - \bm R_j|$.
The parameter $\epsilon$ characterizes the strength of interactions,
and $\sigma=2a$ represents the diameter of particles. 
}
${\bm G_i^V}$ and ${\bm G_i^\Omega}$ are random forces and torques due to thermal fluctuations, which has the following properties  
\begin{eqnarray}
\langle{\bm G^n_i(t)}\rangle=0,\ \ \ \langle{\bm G}_i^n(t){\bm G}_j^n(0)\rangle=\alpha^n {\bm I}\delta(t)\delta_{ij}
\end{eqnarray}
where the square brackets denote taking an average over an equilibrium ensemble. $\alpha^n$ represents the noise intensity for each degree of freedom of the translation {\small($n=V$)} and rotation {\small($n=\Omega$)} of the particles. 
Each noise intensity is controlled so that the variance of the
translational and rotational velocity has a constant value; that is,
$\langle{\bm V_i}^2\rangle = C_1$ and $\langle{\bm \Omega_i^2}\rangle = C_2$, where $C_1$ and
$C_2$ are constant numbers. 
The time evolution of the noise intensity is described by
$\alpha^V(t+\Delta t) =\alpha^V(t)e^{1- \bm V_i^2(t)/C_1}$ and
$\alpha^\Omega(t+\Delta t)=\alpha^\Omega(t)e^{1-\bm \Omega_i^2(t)/C_2}$,
where $\Delta t$ is the discrete time interval which plays a role of
thermostat. 

The temperature of the system is defined by the diffusive motion of the
dispersed particles. 
The translational particle temperature $k_BT^V$ is determined by the
long-time diffusion coefficient $D^V$ of a spherical particle in the
Stokes-Einstein relation $k_BT^V=6\pi\eta a D^V$ where $D^V$ is obtained
from computer simulation. 
Similarly, the rotational particle temperature $k_BT^{\Omega}$ can be
determined by the rotational diffusion coefficient $D^\Omega$. 

There are several advantages to the Langevin approach compared with the
fluctuating hydrodynamics approach, for which TF is introduced in the
Navier-Stokes equation as stochastic stresses to be defined on $N^d$
grid points of fluid simulations.
One important advantage is the computational efficiency: while a
$N^d$ spatial grid requires generating $O$($N^d$) random numbers 
for the fluctuating hydrodynamics, our method requires $O$($N_p$) random
numbers for a dispersion composed of $N_p(\ll  N^d)$ particles. 
Second, if we consider the solvent as a complex fluid--with arbitrary
constitutive equation--, our method has another merit: 
in the fluctuating hydrodynamics approach, the friction tensor for
complex fluids is required, but not here.
Derivations of the friction tensor for complex fluids often have theoretical
difficulties, and even if the friction tensor is obtained, there may be
a large computational cost. 

\section{Test of the Simulation Method}
In order to test our simulation method, we consider the translational
motion of a particle dragged with a constant external force $F_0$ in a
Newtonian fluid. 
With the force turned on, a dragged particle and the solvent have a
steady state solution: the particle has a constant velocity along $x$-direction until $t=0$. Then, at $t>0$, the external force is removed and
the particle and solvent relax toward a rest state due to
dissipation. 

\begin{figure}[h]
\includegraphics[scale=1.]{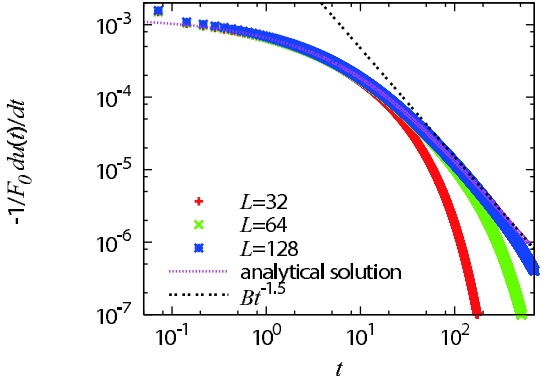}
\caption{\label{RELAX} The response functions, the translational velocity of a dragged particle at a constant external force $F_0$, for system size $32^3$, $64^3$ and $128^3$. 
The response function is affected by the finite-size effect. 
Dotted line is the analytical solution of a drag problem with the time-dependent friction \cite{citeulike:1284488}. 
Dash-dotted line is the algebraic power $Bt^{-3/2}$ where $B=1/12\rho_f(\pi\nu_f)^{3/2}$.}
\end{figure}

Numerical simulation has been performed in three dimensions with
periodic boundary conditions. 
{
The lattice spacing $\Delta$ is taken to be the unit of length.
Other units are defined so that we can set $\eta=1$ and $\rho_f=1$ in
eq.(\ref{nseq}), namely the units of time, mass, pressure are given by 
$\rho_f\Delta^2/\eta$, $\rho_f\Delta^3$, and $\eta^2/\rho_f\Delta^2$, respectively. 
}  
For a spherical particle, the radius $a=5$, the thickness $\xi=2$ and
the particle density $\rho_p=1$ was used.
In Fig.\ref{RELAX}, we plotted $-\dot{u}(t)/F_0$ for system size $32^3$,
$64^3$ and $128^3$ where $u(t)$ denotes the velocity in the $x$-direction of
a dragged particle. 
The response function almost coincides with the analytical solution
based on a time-dependent friction \cite{citeulike:1284488} when the
system size is larger.
The analytical solution is provided in detail in Appendix. 
For the long-time region, the regression of the velocity shows a
power-law decay $Bt^{-3/2}$ where $B=1/12\rho_f(\pi\nu_f)^{3/2}$ and the
kinematic viscosity $\nu_f=\eta/\rho_f$, which depends only on the
hydrodynamic property of the solvent. 

\begin{figure}[h]
\includegraphics[scale=1.]{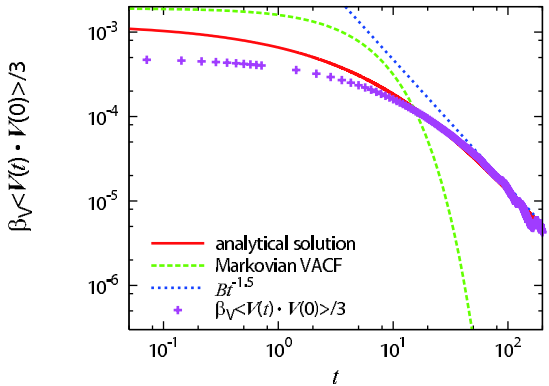}
\caption{\label{VACF} Pluses($+$): Translational velocity autocorrelation function for Brownian particles at $\beta_V=1.2$ and $\Phi=0.002$. The solid line is the analytical solution of a drag problem for the translational particle velocity. The dotted line is the algebraic power $Bt^{-3/2}$ where $B=1/12\rho_f(\pi\nu_f)^{3/2}$. The dashed line is the exponential decay obtained from Markovian Langevin equation.}
\end{figure}

A basic relation between the relaxation response of a dragged particle
and the velocity autocorrelation function of a Brownian particle is the
fluctuation-dissipation theorem (FDT) 
\begin{eqnarray}
-\frac{1}{F_0}\frac{du(t)}{dt} = \frac{\beta_V}{3} \langle  \bm V_i(t)\cdot \bm V_i(0)\rangle
\end{eqnarray}
{where $\langle \bm V_i(t)\cdot \bm V_i(0)\rangle/3$ is the translational velocity autocorrelation
function (VACF) for a Brownian particle and $\beta_V=1/k_BT^V$.
} 
The relation holds that the response function of a dragged particle with
external force $F_0$ is equal to the VACF of a particle in thermal
equilibrium. 
Under the same computational conditions as the relaxation experiment,
the VACF in thermal equilibrium has been calculated at the volume
fraction $\Phi=0.002$ and system size $64^3$. Figure \ref{VACF} shows the
VACF for a Brownian particle. 
We can calculate the long-time diffusion coefficient $D^V$ by the
mean-square displacement of a Brownian particle; that is,
{
$\lim_{t\rightarrow \infty}\langle |\bm R_i(t)- \bm R_i(0)|^2\rangle=6D^Vt$. 
}

When simulating a Brownian particle with the HI, the diffusion
coefficient is affected by finite-size effects and is given by
$D^V=k_BT^V/6\pi\eta a K(\Phi)$ where $K(\Phi)$ represents the effect of
the periodic boundary condition \cite{citeulike:2104188, Hashi}. 
Taking the finite-size correction into account, the diffusion
coefficient at infinite dilution is obtained as $D^V_{inf}=D^V
K(\Phi)$. 
The translational particle temperature is estimated to be
$k_BT^V=6\pi\eta a D^V_{inf} \simeq 0.83$. 
In Fig.\ref{VACF}, $\beta_V\langle \bm V_i(t)\cdot \bm V_i(0)\rangle/3$ approaches to the
analytical solution in the long-time region, shows the power-law decay
B$t^{-3/2}$, and gives the response function of a dragged particle. In
the short-time region, a gap between simulation and the analytical solution is observed. 
The amplitude of a particle's velocity is related to the equipartition
law of energy; that is, $\langle \bm V_i^2\rangle = 3k_BT^V/M_e$ where $M_e$ is the effective
mass. 
The effective mass can be obtained via a hydrodynamic calculation as 
$M_e =M_i +0.5m_0$, where $m_0=4\pi\rho_f a^3 /3$ is the mass of fluid
displaced by a spherical particle with radius a. In this simulation, the
effective mass is estimated to be $M_e \simeq 4.2M_i$, which is notably greater
than the hydrodynamic effective mass. 
The disagreement may be due to the influence of the artificial 
smoothed boundary used between a particle and fluid. 

\begin{figure}[h]
\includegraphics[scale=1.]{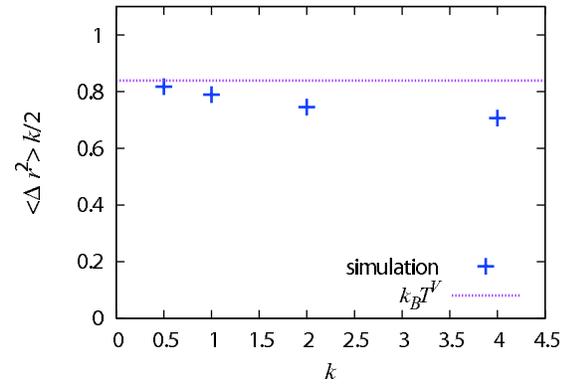}
\caption{\label{spring} The spring constant dependence of the $\lim_{t\rightarrow\infty}\langle \Delta r^2(t) \rangle k/2$. $ k_BT^V$ is the translational particle temperature, which is obtained by the diffusion coefficient for a Brownian particle. }
\end{figure}

Furthermore, we investigate Brownian particles in harmonic potentials. 
{The harmonic potential with a spring constant $k$ is introduced by adding $\bm F^{ex}_i = -k(\bm R_i - \bm R_i^{eq})$, where $\bm R_i^{eq}$ is its equilibrium position, to the equations of motion of particles.}
The mean-square displacement of a Brownian particle trapped in a harmonic potential is given by: 
\begin{eqnarray}
\lim_{t\rightarrow \infty}\langle \Delta r^2(t)\rangle = \frac{2k_BT}{k},
\end{eqnarray}
where $\langle \Delta r^2(t)\rangle =\langle|\bm R_i(t)-\bm R_i(0)|^2\rangle/3$.
In the simulation for the particle number $N_p=8$, each particle is trapped
at the grids of the fcc lattice by harmonic potentials and the direct
interaction between particles is ignored. 
For $k=0.5, 1.0, 2.0$ and $4.0$, the $\lim_{t\rightarrow
\infty}\langle \Delta r^2(t) \rangle k/2$ is plotted in Fig.\ref{spring}. 
The result approaches $k_BT^V$ for $k\rightarrow 0$ and is consistent
with the results obtained from the diffusion coefficient for a Brownian particle. 

\begin{figure}[h]
\includegraphics[scale=1.]{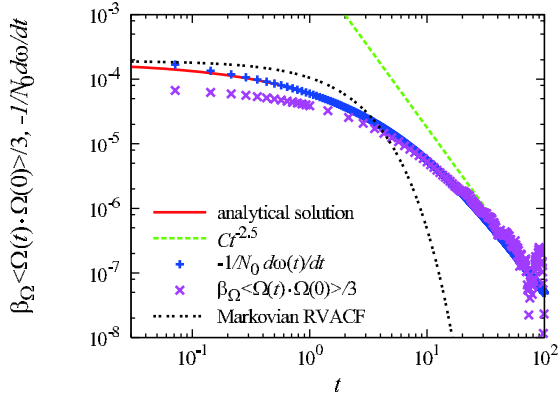}
\caption{\label{RVACF} Crosses($\times $): Rotational velocity autocorrelation function for Brownian particles at $\beta_{\Omega}=1$ and $\Phi=0.002$. Pluses($+$): The relaxation response of rotational velocity of a particle with a constant external torque $N_0$. The solid line is the analytical solution of a drag problem for the rotational particle velocity. The dashed line is the algebraic power $Ct^{-5/2}$ where $C=\pi/32\rho_f(\pi\nu_f)^{5/2}$. The dashed-dotted line is the exponential decay obtained from Markovian Langevin equation.
}
\end{figure}

We investigate the response function of the rotational motion of a
particle with a constant external torque $N_0$ in a similar fashion as
the investigation of the particle's translational motion.
Figure \ref{RVACF} shows the relaxation response $-\dot\omega/N_0$ where
$\omega(t)$ denotes the particle's rotational velocity. 
The response function almost coincides with the analytical solution, and a power-law decay $Ct^{-5/2}$ is observed in the
long-time region where $C=\pi/32\rho_f(\pi\nu_f)^{5/2}$. 
The FDT for the rotational velocity of a particle is also investigated. 
{
The rotational velocity autocorrelation function $\langle \bm \Omega_i(t)\cdot \bm \Omega_i(0)\rangle/3$
(RVACF) is given in Fig.\ref{RVACF}, where $\bm \Omega_i(t)$ denotes the
rotational velocity of a Brownian particle.} 
In the Green-Kubo formula for the rotational velocity of the particle,
the rotational particle temperature is estimated to be 
$k_BT^\Omega =8\pi\eta a^3 D^\Omega \simeq 1$. 
The RVACF also shows a power-law decay $Ct^{-5/2}$ in the long-time region. 
The effective moment of inertia ${\bm I_e}$ can be estimated by the value
of the same time correlation as ${\bm I_e} \simeq 2.6{\bm I_i}$. 

\section{Applications}

As a demonstration of a DNS incorporating thermal fluctuation of
particles, our method is applied to a many-particle system and a dilute
polymeric chain. 

\subsection{Many particles system}

\begin{figure}[h]
\includegraphics[scale=1.]{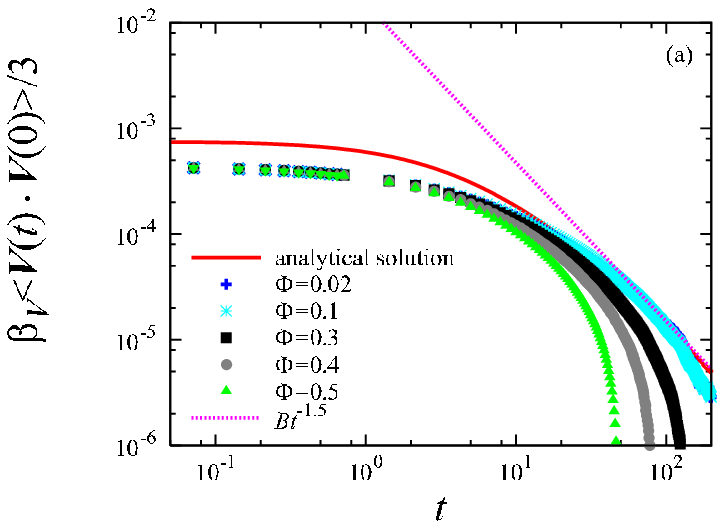}
\includegraphics[scale=1.]{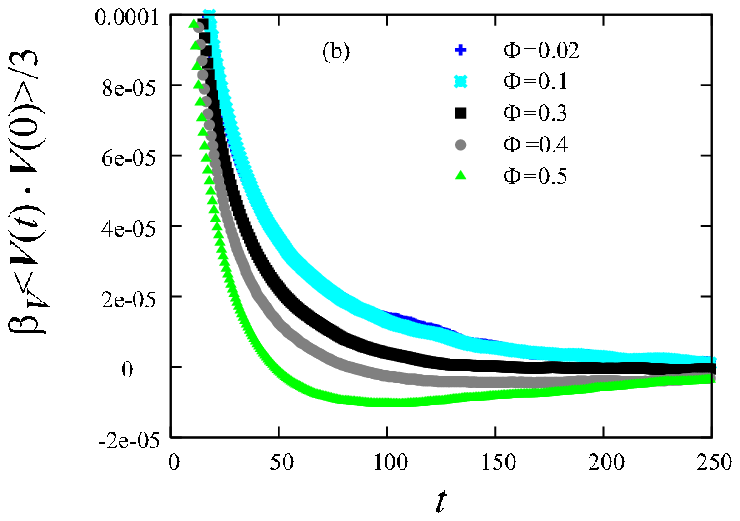}
\caption{\label{MANY_VACF1}
The translational velocity autocorrelation function for each volume fraction. The parameters of the simulation are $\rho_f=1, \eta=1, \rho_p=1, a=5, \xi=2 $, $\beta_V=1$ and $\beta_\Omega=1$. Fig.\ref{MANY_VACF1}(a) is the log-log plot. Fig.\ref{MANY_VACF1}(b) is the normal plot.
}
\end{figure}

In Fig.\ref{MANY_VACF1}, the translational velocity autocorrelation
function for each volume fraction $\Phi$ is presented. 
As the volume fraction is increased, it is found that the relaxation was
more rapid than in low volume fractions, and the power-law long-time
tail of the VACF gradually disappear in Fig.\ref{MANY_VACF1}(a). 
Figure \ref{MANY_VACF1}(b) shows that for $\Phi \geq 0.4$, the VACF has
a negative overshoot, which represents an oscillative motion of a tagged
particle due to a transient cage composed of surrounding particles. 
Compared with Brownian dynamics without HI,
the decay of the correlation for a high volume fraction is much slower
with HI, probably due to the lubrication force between particles.

\subsection{A dilute polymeric chain in good solvent}

The role of the HI is important in the dynamics of a dilute polymeric
chain in a good solvent, and the simple theoretical model is known as
the Zimm model \cite{citeulike:866506}. 
This model treats the HI between beads as a hydrodynamic mobility
matrix, such as the Oseen tensor. 
Some groups have studied a single polymeric chain with the HI using similar
hybrid simulation methods, and their results are in agreement with Zimm theory
\cite{LB,citeulike:2362078, giu}. 
Here, we reexamine the validity of the Zimm model using our present DNS
method since it is supposed to be more accurate than other methods used
previously.

As a model of a polymeric chain, we study a bead-spring model with a
truncated Lennard-Jones potential and {a finitely extensible non-linear elastic (FENE) potential\cite{fene}}: 
\begin{eqnarray}
U_F(r)=-\frac{1}{2}k_cR^2_0\ln\{1-(r/R_0)^2\},
\end{eqnarray}
where $k_c=30\epsilon/\sigma^2$, $R_0=1.5\sigma$ and $r$ is the distance
between the neighboring beads. 
The position vector of a bead is described by ${\bm R_n(t)}$ with 
$n=0,1, . . . ,N_{ch}-1$ where $N_{ch}$ denotes the total number of beads.

\begin{figure}[h]
\includegraphics[scale=1.]{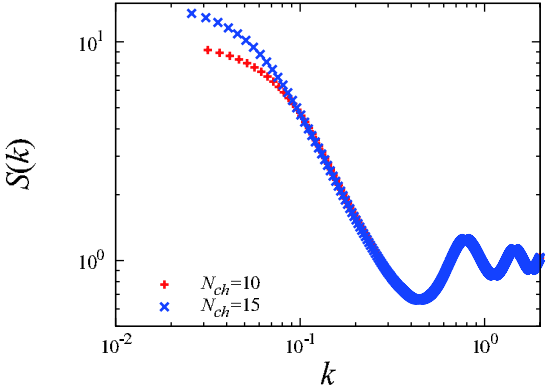}
\caption{\label{Sk} The static structure factor of a chain, for $N_{ch}=10$ (Pluses), $15$ (Crosses). The parameters of the simulation are $\rho_f=1, \eta=1, \rho_p=1, a=5, \xi=2 $, $\beta_V=1$ and $\beta_\Omega=1$. System size is $64^3$ grids.}
\end{figure}

The static property of a polymeric chain is characterized by the static
exponent $\nu$, which is defined as $\langle R_G^2 \rangle\propto N_{ch}^{2\nu}b^2$
for large $N_{ch}$, where $R_G$ is the radius of gyration and $b$ is the
average bond length. 
The static exponent is related to the size of a polymeric chain, which is
$\nu=0.5$ for a Gaussian chain and $\nu\simeq 0.6$ for a self-avoided
chain.
The static exponent $\nu$ of a polymeric chain can be calculated via the
static structure factor $S(k)$. 
Figure \ref{Sk} displays the static structure factor for the bead
numbers $N_{ch} = 10, 15$. 
In the range of $R_G^{-1} \ll k \ll b^{-1}$, $S(k)$ obeys the scaling
relation $S(k) \propto k^{-1/\nu}$ and can determine the static exponent
$\nu \sim 0.62$ by fitting a power law to our data. 

To analyze the relaxation dynamics of a polymeric chain, the real space 
motion is decomposed into a set of the Rouse modes 
($p=0,1,\cdots,N_{ch}-1$), 
\begin{eqnarray}
{\bm X_p}(t)=\frac{1}{N_{ch}}\sum_{n=1}^{N_{ch}}{\bm R_n(t)}\cos[(n-1/2)p\pi/N_{ch}].
\end{eqnarray}
Within the approximations of the Zimm theory, the autocorrelation
function of the Rouse mode decays exponentially as $\langle {\bm
X_p(t)}\cdot{\bm X_p(0)}\rangle /\langle {\bm X_p^2}\rangle = \exp(-t/\tau_p)$, where
$\tau_p$ denotes the relaxation time of the Rouse mode. 
The Zimm theory predicts the relation between the static exponent $\nu$
and the relaxation time $\tau_p$. 
The prediction is $\tau_p \sim p^{-3\nu}$ for the continuous model, 
and $\tau_p \sim p^{2-3\nu}\sin^{-2}(p\pi/2N_{ch})$ for the discrete model.

\begin{figure}[h]
\includegraphics[scale=1.]{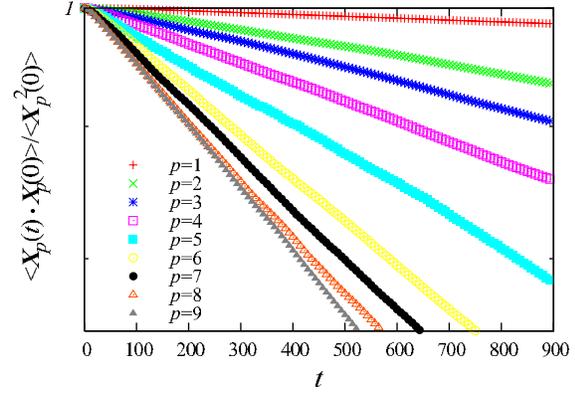}
\caption{\label{Xp} Normalized autocorrelation functions of the Rouse mode $X_p$ for various mode p. The chain length is $N_{ch}=10$.}
\end{figure}

\begin{figure}[h]
\includegraphics[scale=1.]{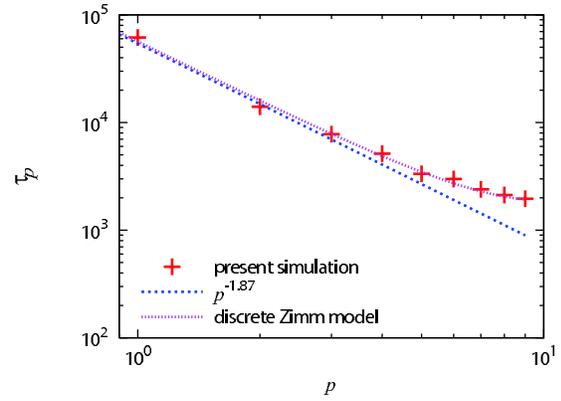}
\caption{\label{TIME} Rouse mode $p$-dependence of the relaxation time for chain length $N_{ch}=10$. The dashed line is a fitting power law to simulation data for $p\leq 5$. The theoretical prediction of the discrete Zimm model is indicated as a dotted line, $\tau_p \sim p^{2-3\nu}\sin^{-2}(p\pi/2N_{ch})$.}
\end{figure}

Figure \ref{Xp} shows the normalized autocorrelation function of the
Rouse mode for $N_{ch}=10$. 
The relaxation times are obtained by a fitting in the exponential
short-time regime $t \in [50:1000]$. 
Figure \ref{TIME} shows the mode ($p$-) dependence of the relaxation time. 
By fitting for $p\leq 5$, the $p$-dependence of the relaxation time is
estimated as $\tau_p \sim p^{-1.87}$. 
In Fig. \ref{TIME}, the prediction of the discrete Zimm model for the
$p$-dependence of $\tau_p$ is also plotted using $\nu$ obtained from
$S(k)$. 
The numerical results show a good agreement with the prediction of 
the Zimm model. 

\section{Summary}

We have developed a numerical method for consistently implementing
thermal fluctuations and hydrodynamic interactions into models of the
motions of Brownian particles dispersed in incompressible host fluids. 
We represented the thermal fluctuations by random forces acting on
Brownian particles and the hydrodynamic interactions by directly
resolving the fluid motions.
The validity of the method has been examined carefully by comparing the
present numerical results with the fluctuation-dissipation theorem for a
dispersion of a single spherical particle. 
Simulations are then performed for dispersions of many spherical
particles, and also for a polymeric chain in a fluid. 
In the former case, we found that the hydrodynamic long-time tail in the
VACF--clearly observed for a single particle dispersion-becomes weak with
increasing volume fraction of the particles. 
In the latter case, we found that our numerical results coincide quite
well with the theoretical predictions of the Zimm model. 

\appendix
{
\section{A Dragged Particle with the Time-dependent Friction}

From the hydrodynamics, we can obtain the time-dependent friction\cite{citeulike:1284488}
\begin{eqnarray}
\zeta(t) = -\Bigl[6\pi\eta a u(t) + \frac{2}{3}\pi\rho_fa^3\dot u(t) + 6 a^2 \sqrt{\pi\eta\rho_f}\int_{-\infty}^t ds \frac{\dot u(s)}{\sqrt{t-s}}\Bigl].
\end{eqnarray}
The first term is the standard Stokes resistance. 
The second term represents the additional mass, which is related to the
acceleration of the particle. 
The third term represents the memory effect, which addresses the
temporal decay of the fluid's momentum. 

The equation of motion of a dragged particle with a time-dependent
frictional force $\zeta$ and a constant external force $F_0$ is
described by $M \dot u(t) = \zeta(t) + F_0$, where M is the mass of a
particle. 
After the particle has reached a steady state (a constant velocity), the
external force is removed. 
Then, the regression of the velocity can be written as 
\begin{eqnarray}
\frac{du(t)}{dt} = - \frac{F_0}{M_{eff}}\int_0^{\infty} \frac{dy}{\pi}
 \frac{\sigma_0 \sqrt{y}e^{-y\frac{|t|}{\tau_B}}}{|1-y|^2 + \sigma_0^{2}
 y}\label{TRANS_ANA}, 
\end{eqnarray}
where $\sigma_0 = (9\rho_f/(2\rho_p + \rho_f))^{1/2} $, $M_{eff}= M+0.5m_0$, $\rho_p$ the
particle density and $\tau_B=M/6\pi\eta a$.
From the fluctuation-dissipation relation $\frac{1}{3k_BT^V}\langle \bm V_i(t)\cdot\bm V_i(0)\rangle = -
\frac{1}{F_0}du(t)/dt$, a hydrodynamic velocity autocorrelation function
of a Brownian particle can be calculated  exactly. 

Similarly, the regression of the rotational velocity $\omega$ of a
spherical particle with a moment of inertia $I$ can be written as 
\begin{eqnarray}
&&\frac{d\omega(t)}{dt}=,\\
&&-\frac{N_0}{8\pi\eta a^3}\frac{1}{\tau_f}\int_0^{\infty}\frac{dy}{3\pi} \exp(-yt/\tau_f)
\Biggl[ \frac{y^{3/2}}{[1-(\frac{\tau_r}{\tau_f} + \frac{1}{3}) y]^2 + y (1 - \frac{\tau_r}{\tau_f}y)^2 }\Biggl] \label{ROT_ANA},
\end{eqnarray}
where $N_0$ is a constant external torque, $\tau_f=a^2/\nu_f$ and
$\tau_r=I/8\pi\eta a^3$. 
}



\end{document}